\documentstyle[12pt]{article}
\date{\nonumber}
\begin{document}
\setlength{\baselineskip}{16.5pt}
\noindent {\large Letter}

\vskip 6mm
\centerline{
{\Large Virasoro Symmetry Algebra of Dirac Soliton Hierarchy}}

\vskip 3mm
\centerline{{\large Wen-Xiu Ma$^{\dagger \ddagger}$\footnote{Email: 
wenxiuma@uni-paderborn.de}
 and Kam-Shun Li$^{\S}$}}

\vskip 1mm
\centerline{{\small$^{\dagger}$Institute of 
Mathematics, Fudan University, Shanghai 200433, P. R. China }}

\centerline{{\small
$^{\ddagger}$FB Mathematik-Informatik, Universit\"at-GH Paderborn,
D-33098 Paderborn, Germany}}

\centerline{{\small
$^{\S}$Department of Mathematics, Chinese University of Hong Kong, Shatin,
N.T., Hong Kong}}

\begin{abstract} 
A hierarchy of first-degree time-dependent 
symmetries is proposed for Dirac soliton
hierarchy and their commutator relations with time-dependent symmetries 
are exhibited. Meantime,
a hereditary structure of Dirac soliton hierarchy is elucidated
and a Lax operator algebra associated with Virasoro  symmetry algebra
is given.
\end{abstract}

\def\be{\begin{equation}}
\def\ee{\end{equation}}
\def\ba{\begin{array}}
\def\ea{\end{array}}
\def\la {\lambda}
\def \part {\partial}
\def \al {\alpha}
\def \de {\delta}
\def \si {\sigma}

\vskip 0.5cm
The main purpose of the present letter is to construct a hierarchy of
first-degree time-dependent symmetries for Dirac soliton hierarchy 
\cite{Grosse}.
This kind of symmetries was first proposed in the famous Olver's paper 
\cite{Olver}. They include ones corresponding to Galilean invariance 
and scalar invariance and  
they are generally related to the first-degree
 master symmetries \cite{Fuchssteiner}.
Afterwards, 
some theory to describe time-dependent 
symmetries were developed for various
classes of soliton equations \cite{ChenLLetal} \cite{OevelFM}. 
Moreover it is found that for 
some nonlinear systems, there exist $W_\infty$ symmetry algebras 
involving arbitrary functions of some independent variables, for example,
time variable $t$ \cite{Winternitzetal}. However,
for systems of evolution equations, 
 the kind of symmetries involving
 arbitrary functions of time variable $t$ doesn't exist
\cite{Ma1991}.
These systems may possess polynomial time-dependent symmetries, which relate
to master symmetries of any degree.
 Usually only first-degree time-dependent symmetries 
can be  found for soliton equations in $1+1$ dimensions.  
Recently,
in terms of Lax operators, a simple but systematic scheme for generating
first-degree time-dependent symmetries in $1+1$ dimensions
has been established in \cite{Ma1993a}.
Here we would like to discuss the case of Dirac soliton
hierarchy through that trick.

Dirac soliton hierarchy reads as \cite{Grosse} \cite{Ma1996}
\be u_{t_m}=
\left(\ba {l}  q\\r\end{array} \right) _{t_m}=K_m=\Phi ^m f_0=\Phi ^m J
f_0'=J\Psi ^mf_0'=J
\frac {\delta H_m}{\delta u}
,\ m\ge 0,\label{Dirac}\ee 
with 
\be J=\left(\ba {cc}  0&-1\\ 
\vspace {1mm}
1&0\end{array} \right) ,\ 
\Phi = \Psi ^*=\left(\ba {cc}  2r\part ^{-1} q & -\frac12 
\part +2r\part ^{-1} r\\
\vspace {1mm}
\frac12 \part -2q\part ^{-1} q & -2q\part ^{-1} r\end{array} \right) ,\ f_0=
\left(\ba {c}  
2r\\
\vspace {1mm}
-2q\end{array} \right) .\label{hereo}\ee
Here $\part =\part /\part x$ and $\part ^{-1}\part =\part \part ^{-1}=1.$
After this hierarchy was presented by Grosse \cite{Grosse}, 
it did not arouse enough attention until
its Hamiltonian structure and binary nonlinearization
were recently established (e.g. see \cite{Ma1996}).
It is not difficult to find that $J$ and $M=\Phi J$ constitute a pair of 
Hamiltonian operators and thus Dirac soliton hierarchy (\ref{Dirac}) possesses
a bi-Hamiltonian structure. It differs 
from Kaup-Newell hierarchy but assembles AKNS hierarchy.
If we move the derivative $\part $
from off-diagonal to diagonal and interchange the 
positions of two potentials in $\Phi$,
the recursion operator $\Phi$ of Dirac 
hierarchy will be transformed into one of AKNS hierarchy.
The first nonlinear Dirac system in Dirac hierarchy (\ref{Dirac}) is as
 follows 
\[ \left \{ \begin{array}{l}q_{t_2}=-\frac 12 r_{xx}+q^2r+r^3,\vspace{1mm}\\
 r_{t_2}=\frac 12 q_{xx}-q^3-qr^2.\end{array} \right.\]
This system is different from the coupled nonlinear Schr\"odinger
system in AKNS hierarchy because it contains the cubic terms
$q^3,\,r^3$.

Dirac soliton hierarchy (\ref{Dirac}) 
associates with the following spectral problem
\be \phi _x=U\phi =U(u,\la )\phi 
,\ \phi =
\left(\ba {l} \phi _1 \\ 
\vspace {1mm}
\phi _2 \ea \right) ,\ 
U= \left(\ba {cc}  q & \la + r \\ 
\vspace {1mm}
-\la + r  &- q  \ea \right).
\label{sp} \ee 
There have been many results about this Dirac spectral problem.
A Gelfand-Levitan-Marchenko equation holds for  Dirac spectral 
problem (\ref{sp})
\cite{Frolov}. Time evolution of scattering data of (\ref{sp})  was discussed 
in  \cite{Grosse} and a detailed analysis on inverse scattering 
problem was provided by Hinton et al. \cite{HintonJKS} for a more general
spectral problem.
Its trace formula has also been carefully investigated in \cite{Shi}.
Its isospectral ($\la _{t_m}=0$) flows are exactly Dirac soliton hierarchy
(\ref{Dirac}).
We would like to derive the  nonisospectral 
 $(\la _{t_n}=\la ^n,\ n\ge0)$
Dirac flows corresponding to (\ref{sp}) and then 
present a Virasoro symmetry algebra
of (\ref{Dirac}). 

In order  to take advantage of the trick in \cite{Ma1993a}, we need to solve 
the characteristic operator equation with respect to $\Omega=\Omega (X)$:
\be [\Omega ,U]+\Omega _x =U'[\Phi X]-\la U'[X]\label{cq}\ee 
with any fixed vector field $X=(X_1,X_2)^T$
and an initial nonisospectral $(\la _{t_0}=1)$ key equation with
respect to $g_0,\,B_0$:
\be U'[g_0]+U_\la -B_{0x}+[U,B_0]=0.\label{kq}\ee
It is easy to work out 
\[U'[X]=\left(\ba {cc} 
X_1&X_2\vspace {1mm}\\
X_2&-X_1\ea \right),\ \Phi X=\left(\ba {c} 
-\frac12 X_{2x}+2r\part ^{-1}(qX_1+rX_2)\vspace {1mm}\\ 
\frac12 X_{1x}-2q\part
^{-1}(qX_1+rX_2)\ea \right).\]
It follows that
\be \ba {l}\quad 
U'[\Phi X]-\la U'[X]
\\=
\left(\ba {cc} 
-\frac12 X_{2x}+2r\part ^{-1}(qX_1+rX_2)
&
\frac12 X_{1x}-2q\part ^{-1}(qX_1+rX_2)\vspace {1mm}\\
 \frac12 X_{1x}-2q\part
^{-1}(qX_1+rX_2)
&
 \frac12 X_{2x}-2r\part
^{-1}(qX_1+rX_2)
\ea \right)-\la \left(\ba {cc} 
X_1&X_2\vspace {1mm}\\
X_2&-X_1\ea \right).\ea \label{cqr}\ee 
If we choose a special form of $\Omega$
\[ \Omega =\left(\ba {cc}  c&a+b\\
\vspace {1mm}
a-b&-c\ea \right)=a\si _1+bi\si _2+c\si _3,\ i=\sqrt{-1}\,,\]
where $\si _i,\ 1\le i\le 3,$ are Pauli $2\times 2$ matrices, then we have
\be [\Omega ,U]+\Omega _x=\left (\ba {cc}  -2\la a+2rb+c_x & 2\la
c-2qb+2rc-2qa+a_x+b_x\vspace {1mm}\\ 
2\la c-2qb-2rc+2qa+a_x-b_x & 2\la a -2rb-c_x\ea \right ). \label{cql}\ee
The substitution of (\ref{cqr}) and (\ref{cql}) into the 
characteristic operator equation (\ref{cq})
results in 
$$a=\frac12 X_1,\ b=\part ^{-1}(qX_1+rX_2),\ c=-\frac12 X_2.$$
Thus we obtain an operator solution to 
the characteristic operator equation (\ref{cq})
\be \Omega =\Omega (X)
=\left (\ba {cc}  -\frac12 X_2&\frac12 X_1+\part ^{-1}(qX_1+rX_2)\vspace {1mm}
\\
\frac12 X_1-\part ^{-1}(qX_1+rX_2)&\frac12 X_2\ea \right ).\label{cs}\ee
By a similar argument, we may find a pair of solutions to 
the initial key equation (\ref{kq})
\be B_0=\left (\ba {cc}  0&x\vspace {1mm} \cr
-x &0\ea \right ),\ g_0=\left (\ba {c}  2xr\vspace {1mm}
\cr
-2xq\ea \right ).\label{ks}\ee 

The above results allow us to conclude
that 
\be u_{t_n}=\rho  _ n =\Phi ^n g_0,\ n\ge0\label{nDirac}\ee
is just the required 
hierarchy of nonisospectral $(\la _{t_n}=\la ^n,\ n\ge0)$ flows
and this hierarchy possesses zero curvature representations
\be U_{t_n}-W_{nx}+[U,W_n]=0,\ \textrm{i.e.}\ 
U'[\rho  _n]+\la ^n U_\la -W_{nx}+[U,W_n]=0\label{nLax}\ee 
with the spectral evolution laws 
$\la _{t_n}=\la ^n,\ n\ge0,$
and Lax operators
\be W_{n}=\sum_{j=0}^n\la ^{n-j}B_j=\la ^nB_0+\sum_{j=1}^n\la
^{n-j}\Omega (\rho  _{j-1}).\label{nLaxo}\ee 
In fact, we can calculate that
\[ \ba {l}\quad W_{nx}-[U,W_n]\vspace{1mm}\\
=\sum_{j=0}^n\la ^{n-j}(B_{jx}-[U,B_j])\vspace{1mm}\\
=\la ^nU'[g_0]+\la ^nU_\la +\sum_{j=1}^n(U'[\rho _j]-\la U'[\rho _{j-1}])
\vspace{1mm}\\ =\la ^nU_\la +U'[\rho _n],\ n\ge 1, \ea 
\]
and thus zero curvature equations (\ref{nLax}) for $n\ge 0$ hold.
Completely similar to the deduction of the nonisospectral case \cite{Ma1993b},
we can obtain the following isospectral $(\la _{t_m}=0)$ 
Lax operators associated with Dirac
soliton hierarchy (\ref{Dirac})
\be V_{m}=\sum_{i=0}^m\la ^{m-i}A_i=\la ^mA_0+\sum_{i=1}^m\la
^{m-i}\Omega (K _{i-1}),\ A_0=\left (\ba {cc}  0&1\\
\vspace{1mm} -1&0\ea \right ).\label{Laxo}\ee 
It is known that 
IST technique can be applied to not only isospectral soliton equations,
but also  nonisospectral ones
(e.g. see \cite{ChanL}).
The above neat forms of isospectral and nonisospectral Lax operators
may provide us with some help to carry out 
IST technique. The nonisospectral Lax operators $V_m,\ m\ge 0$, are all
local. But the nonisospectral Lax operators $W_n,\ n\ge 0$, are nonlocal
except the first two ones $W_0, W_1$ since $\rho _0=(2xr,-2xq)^T,\ 
\rho _1=(q+xq_x,r+xr_x)^T$ are local and 
the other nonisospectral vector fields $\rho _n,\ n\ge 2,$ are
nonlocal.

In what follows, we want to show that 
the nonisospectral flows (\ref{nDirac}) are all the first-degree 
master symmetries of Dirac hierarchy (\ref{Dirac}) and simultaneously to
establish a Virasoro symmetry algebra for Dirac
hierarchy (\ref{Dirac}). 

Towards this end,
first we directly prove the operator $\Phi$ in (\ref{hereo}) is a
hereditary operator \cite{Fuchssteiner1979}, 
which also shows that Dirac soliton hierarchy possesses a kind of nice
structure, i.e.
hereditary structure. 
The hereditariness of $\Phi$ means 
that $\Phi$ satisfies the following equation
\be \Phi '[\Phi X]Y-\Phi '[\Phi Y]X-
\Phi \{\Phi'[ X]Y-\Phi '[ Y]X\}=0\label{hp}\ee
for any two vector fields $X,Y$, which is exactly the same as
an operator identity that
 is proposed for analyzing the reason why soliton equations
come in hierarchies in
\cite{Yordanov}. For ease of writing, 
we introduce an equivalent relation:
\[K\cong S\iff (K-S)(X,Y)=(K-S)(Y,X)\]
for two expressions $K(X,Y),\,S(X,Y)$ depending on vector fields
$X,Y$.
By this 
equivalent relation,
the equality (\ref{hp}) becomes
\[ \ba {l}
\Phi '[\Phi X]Y-\Phi \Phi'[ X]Y\vspace{2mm}
\\=(\cdots,
(\Phi '[\Phi X]Y)_i,\cdots)^T-
(\cdots,
(\Phi \Phi'[ X]Y)_i,\cdots)^T\cong 0.\ea \]
Let $P=qX_1+rX_2,\,
Q=qY_1+rY_2$ for $X=(X_1,X_2)^T,\, Y=(Y_1,Y_2)^T$.
A direct calculation can give rise to
\[\ba {l}
\Phi '[\Phi X]=\left (\ba {c} 
[X_{1x}-4q(\part ^{-1}P)]\part ^{-1}q +r\part ^{-1}[-X_{2x}+4r(\part ^{-1}P)],
\vspace{2mm}\cr
[X_{2x}-4r(\part ^{-1}P)]\part ^{-1}q +q\part ^{-1}[X_{2x}-4r(\part ^{-1}P)],
\ea \right.
\vspace{2mm} \cr
\quad\qquad\qquad 
\left.\ba {c}
[X_{1x}-4q(\part ^{-1}P)]\part ^{-1}r +r\part ^{-1}[X_{1x}-4q(\part ^{-1}P)]
\vspace{2mm}\cr
[X_{2x}-4r(\part ^{-1}P)]\part ^{-1}r +q\part ^{-1}[-X_{1x}+4q(\part ^{-1}P)]
\ea \right),\vspace{2mm}\cr
\Phi '[X]Y\cong
\left (\ba {c} 2X_2(\part ^{-1}Q)
\vspace{2mm}\cr
- 2X_1(\part ^{-1}Q)\ea \right ).\ea \]
Further one may acquire
\begin{eqnarray}&& (\Phi'[\Phi X]Y)_1=
[X_{1x}-4q(\part ^{-1}P)](\part ^{-1}Q)\nonumber \\
&&\qquad \qquad  +r\part ^{-1}(X_{1x}Y_2-X_{2x}Y_1)
+4r\part ^{-1}
[(rY_1-qY_2)(\part ^{-1}P)]
,\label{e1}\\ &&
(\Phi'[\Phi X]Y)_2=
[X_{2x}-4r(\part ^{-1}P)](\part ^{-1}Q)\nonumber \\
&&\qquad \qquad   +q\part^{-1}(X_{2x}Y_1-X_{1x}Y_2)+4q
\part ^{-1}[(qY_2-rY_1)(\part ^{-1}P)]
\label{e2};\\
&&(\Phi\Phi'[ X]Y)_1\cong 
4r\part ^{-1}[(qX_2-rX_1)(\part ^{-1}Q)]+[X_1(\part ^{-1}Q)]_x
,\label{e3}\\ &&
(\Phi\Phi'[ X]Y)_2\cong 
-4q\part ^{-1}[(qX_2-rX_1)(\part ^{-1}Q)]+[X_2(\part ^{-1}Q)]_x
.\label{e4}\end{eqnarray}
Therefore by (\ref{e1}) and (\ref{e3}), one has 
\[ \ba {l}
(\Phi\Phi'[ X]Y)_1-(\Phi'[\Phi X]Y)_1\vspace{2mm}\\ \cong 
4r\{\part ^{-1}[(qX_2-rX_1)(\part ^{-1}Q)]-\part ^{-1}[(rY_1-qY_2)(\part ^{-1}
P)]\}\vspace{2mm}\\
+\{[X_1(\part ^{-1}Q)]_x-X_{1x}(\part ^{-1}Q)-r\part ^{-1}(X_{1x}Y_2
-X_{2x}Y_1)\}
\vspace{2mm}\\
\cong  \{
[X_1(\part ^{-1}Q)]_x-X_{1x}(\part ^{-1}Q)-r\part ^{-1}(X_{1x}Y_2-X_{2x}Y_1)\}
\cong 0.\ea \]
With the same argument by (\ref{e2}) and (\ref{e4}), one can verify that
\[
(\Phi\Phi'[ X]Y)_2-(\Phi'[\Phi X]Y)_2\cong 0.\]
Therefore $\Phi$ defined by (\ref{hereo}) is a hereditary symmetry, indeed.
We may also show that the operator $\Phi(\al )$
 with arbitrary constant coefficient
$\al $ of $\part $
\[\Phi (\al ) =\left(\ba {cc}  2r\part ^{-1} q & -\al 
\part +2r\part ^{-1} r\vspace{1mm}
\\
\al \part -2q\part ^{-1} q & -2q\part ^{-1} r\end{array} \right) 
:=\left(\ba {cc}0 & -\al \vspace{1mm}
\cr \al &0  \end{array} \right)\part  +\Phi_0
\] 
is still hereditary and that there exists only this sort of hereditary 
operators among $\Lambda \part +\Phi _0$, where $\Lambda $ is any constant 
matrix.

Secondly, we need the following relation 
on the triple $(\Phi,
f_0,g_0)$, which 
may directly be shown,
\be L_{f_0}\Phi=0,\ 
L_{g_0}\Phi=1,\ \Phi[f_0,g_0]=[f_0,\Phi g_0]=0,\label{rel}\ee
where the Lie derivative $L_X\Phi$ with respect to $X$ is defined by 
\be L_X\Phi =\Phi'[X]-[X',\Phi]=\Phi'[X]-X'\Phi+\Phi X'\label{Lied}\ee
and the commutator $[X,Y]$ of two vector fields,
by $[X,Y]=X'[Y]-Y'[X].$

In view of the known results given in \cite{OevelFM} in the case of 
(\ref{hp}) and (\ref{rel}),
one easily obtains a vector field Lie algebra
\be \left \{\begin{array} {l}
[K_m,K_n]=[\Phi ^m f_0,\Phi ^n f_0]=0,\ m,n\ge 0,
\vspace{2mm}\cr  
[K_m,\rho  _n]=[\Phi ^m f_0,\Phi ^n g_0]=mK_{m+n-1},\ K_{-1}=0,\ m,n\ge 0,
\vspace{2mm}\cr 
[\rho  _m,\rho  _n]=
[\Phi ^m g_0,\Phi ^n g_0]=(m-n)\rho  _{m+n-1},\ \rho  _{-1}=0,
\ m,n\ge 0. \end{array}\right. \label{vfLiea}\ee
For example, one can also directly show that
\[\ba {l}
[K_m,\rho  _n]=[\Phi ^m f_0,\Phi ^n g_0]
\vspace{2mm}\\ 
=\Phi ^n[\Phi ^mf_0,g_0]\ \ (\textrm{due to $L_{f_0}\Phi=0$})
\vspace{2mm}\\
=\Phi ^{n+1}[\Phi ^{m-1}f_0,g_0]+\Phi ^{m+n-1}f_0
\vspace{2mm}\\
=\Phi ^{n+2}[\Phi ^{m-2}f_0,g_0]+2\Phi ^{m+n-1}f_0
\vspace{2mm}\\
=\cdots\cdots
\vspace{2mm}\\
=\Phi ^{m+n}[f_0,g_0]+m\Phi ^{m+n-1}f_0
\vspace{2mm}\\
=mK_{m+n-1}.\ea \]
Here from the third step, we have used the equality $[X, \Phi Y]=\Phi [X,Y]-(
L_X\Phi)Y$, induced by (\ref{Lied}), $m$ times.
Those relations imply that the vector fields $\rho  _n,\ n\ge0$, are all
common master symmetries of the first degree for the whole hierarchy 
(\ref{Dirac}) and
thus one sees that the $l$th Dirac system $u_{t_l}=K_l$ in Dirac
soliton hierarchy (\ref{Dirac}) has
infinitely many first-degree time-dependent symmetries 
\be \tau _n^{(l)}=t[K_l,\rho  _n]+\rho 
_n=ltK_{n+l-1}+\rho  _n,\ n\ge0,\ee 
besides infinitely many time-independent symmetries $K_m, m\ge0$.
The symmetries $\tau _0^{(l)}$ and $\tau _1^{(l)}$ correspond to Galilean
transformation group and scalar transformation group, repectively.
Furthermore these two hierarchies of symmetries constitute a semi-product of a
Kac-Moody algebra and a centerless Virasoro algebra. More precisely, they
possess the following commutator relations
\be \left \{\ba {l} [K_m,K_n]=0,\ m,n\ge 0,
\vspace{2mm}\cr 
[K_m,\tau _n^{(l)}]=mK_{m+n-1},\ m,n\ge 0,
\vspace{2mm}\cr  
[\tau _m^{(l)},\tau _n^{(l)}]=(m-n)\tau _{m+n-1}^{(l)},\ \tau _{-1}^{(l)}=0,
\ m,n\ge 0.\ea \right.\label{Virasoroa}\ee
It is referred to as a Virasoro symmetry 
algebra or hereditary algebra \cite{Fuchssteiner1990} of symmetries.

The above symmetry algebraic structure may also be derived from 
a Lax operator algebra
of $V_m,\,W_n$ determined  by (\ref{Laxo}) and (\ref{nLaxo})
\be \left \{\begin{array} {l}
\lbrack\!\lbrack V_m,V_n
\rbrack\!\rbrack
:=V_m {}' [K_n]-V_n {}'[K_m]+[V_m,V_n]=0
,\ m,n\ge 0,\vspace{2mm}\cr
\lbrack\!\lbrack V_m,W_n
\rbrack\!\rbrack
:=V_m {}' [\rho  _n]-W_n {}'[K_m]+[V_m,W_n]+\la ^nV_{m\la
}\vspace{2mm}\cr
\qquad \quad\ \,\quad =mV_{m+n-1},\ V_{-1}=0
,\ m,n\ge 0,\vspace{2mm}\cr  \lbrack\!\lbrack W_m,W_n \rbrack\!\rbrack
:=W_m {}' [\rho  _n]-W_n {}'[\rho  _m]+[W_m,W_n]+\la ^nW_{m\la
}-\la ^mW_{n\la }\vspace{2mm}\cr
\qquad \quad \ \,\quad\ =(m-n)W_{m+n-1},\ W_{-1}=0 ,\ m,n\ge 0,\ea \right.\ee 
which can be given in a similar way to \cite{Ma1993c} or
by some direct deduction.
A similar Lie algebraic structure for AKNS hierarchy has been shown 
in \cite{ChenZM}.
We have known that if the equalities 
\[\ba {l} U'[K]+f(\la )U_\la -V_x+[U,V]=0,\vspace {1mm}\\
U'[S]+g(\la )U_\la -W_x+[U,W]=0\ea \]
hold, then we have \cite{Ma1993c} 
\[ U'[\,[K,S]\,]+ \lbrack\!\lbrack f,g\rbrack\!\rbrack U_\la -
\lbrack\!\lbrack V,W\rbrack\!\rbrack_x+[U,\lbrack\!\lbrack V,W\rbrack\!\rbrack]
=0,\]
where $[K,S]$ is a commutator of $K,S$, and 
$\lbrack\!\lbrack f,g\rbrack\!\rbrack$ and 
$\lbrack\!\lbrack V, W\rbrack\!\rbrack$ are defined by 
\[\lbrack\!\lbrack f,g\rbrack\!\rbrack(\la )
=f'(\la )g(\la )-f(\la )g'(\la ),\ 
\lbrack\!\lbrack V, W\rbrack\!\rbrack=V'[S]-W'[K]+[V,W]+gV_\la -fW_\la .\]
By using this result and the injective 
property of the Gateaux derivative operator $U':K\mapsto U'[K]$, 
we right now obtain the vector field Lie algebra (\ref{vfLiea}) and further 
Virasoro symmetry algebra
(\ref{Virasoroa}) from the above Lax operator algebra.

To summarize, we have constructed a hierarchy of first-degree time-dependent 
symmetries and have given the commutator relations of the resulting 
time-dependent symmetries and the original time-independent symmetries.
This kind of symmetry algebras is also a common property enjoyed by soliton 
equations. 

\vskip 0.5cm
\noindent {\bf 
Acknowledgments:} 
The authors acknowledge financial support
by an Earmarked Grant for Research of Hong Kong,
the National Natural Science Foundation of China
and the Shanghai Science and Technology Committee of China. 
One (W. X. Ma) of the authors 
would also like to thank the 
Alexander von Humboldt Foundation for a research fellowship
and to thank 
Prof. B. Fuchssteiner and Dr. W. Oevel for valuable discussions.

\small 
\setlength{\baselineskip}{12pt}


\begin{thebibliography}{99}
\bibitem{Grosse} Grosse H 1986 {\it Phys. Rep.} {\bf 134} 297
\bibitem{Olver} Olver P J 1980 {\it Math. Proc. Camb. Phil. Soc.} {\bf
 88}  71
\bibitem{Fuchssteiner} Fuchssteiner B 1983 {\it Prog. Theor. Phys.} {\bf 70}
1508
\bibitem{ChenLLetal} Chen H H, Lee Y C and Lin J E 1982 {\it Advances in 
Nonlinear Waves} ed Debnath L (Boston: Pitman) p233;
Li Y S and Zhu G C 1986 {\it J. Phys. A: Math. Gen.} {\bf 19} 3713
\bibitem{OevelFM} Oevel W 1987 {\it Topics in Soliton Theory and Exactly
Solvable Nonlinear Equations} eds Ablowitz M, Fuchssteiner B and
Kruskal M (Singapore: World Scientific) p108;
Fokas A S  1987 {\it Stud. Appl. Math.} {\bf  77}  253;
Ma W X 1990 {\it J. Phys. A: Math. Gen.} {\bf 23} 2707;
Lou S Y and Chen W Z 1993 {\it Phys. Lett. A} {\bf  179}  271
\bibitem{Winternitzetal} 
  Schwarz F 1982 {\it J. Phys. Soc. Jpn.} {\bf 51} 2387;
  David D, Kamran N, Levi D and Winternitz P 1985 {\it Phys. Rev. Lett.}
 {\bf 55} 2111; Champagne B and Winternitz P 1988
  {\it  J. Math. Phys.} {\bf 29} 1;
  Martina L and Winternitz P 1989 {\it Ann. Phys.} {\bf 196} 231;
  Rubin J and Winternitz P 1990 {\it J. Math. Phys.} {\bf 31}
  2085;
  Lou S Y, Yu J and Lin J 1995 {\it J. Phys. A: Math. Gen.} {\bf 28} L191
\bibitem{Ma1991} Ma W X 1991 {\it Science in China A} {\bf 34} 769
\bibitem{Ma1993a} Ma W X 1993 {\it Phys. Lett. A} {\bf 179} 179; 
1993 {\it Proceedings of the 21st International Conference on the
      Differential Geometry Methods in Theoretical Physics} ed Ge M L
(Singapore: 
World Scientific) p535; 1992
{\it J. Phys. A: Math. Gen.} {\bf 25} L719
\bibitem{Ma1996} Ma W X 1996 Binary nonlinearization 
for Dirac integrable hierarchy, to appear in {\it Chinese Annals of Math. B}
(solv-int/9512002)
\bibitem{Frolov} Frolov I S 1972 {\it Soviet Math. Dolk.} {\bf 13} 1468
\bibitem{HintonJKS} Hinton D B, Jordan A K, Klaus M and Show J K 1991
{\it J. Math. Phys.} {\bf 32} 3015
\bibitem{Shi} Shi Q C 1993 {\it Acta Mathematica Scientia} {\bf 13} 316
\bibitem{Ma1993b} Ma W X 1993 {\it Chinese Science Bulletin} {\bf 38}
2025 
\bibitem{ChanL} Chan W L and Zheng Y K 1987 
{\it Lett. Math. Phys.} {\bf 14} 293;
Chan W L and Li K S 1989 {\it J. Math. Phys.} {\bf 30} 2521; 
Chan W L and Li K S 1994
{\it J. Phys. A: Math. Gen.} {\bf 27} 883
\bibitem{Fuchssteiner1979} 
Fuchssteiner B 1979 {\it Nonlinear Anal. Theor. Meth. Appl.} {\bf 3} 849
\bibitem{Yordanov}
Yordanov R G 1993 {\it J. Math. Phys.} {\bf 34} 4045
\bibitem{Fuchssteiner1990}
Fuchssteiner B 1990 
{\it Nonlinear Dynamics} (Research Reports in
Physics) (Berlin: Springer-Verlag) p114
\bibitem{Ma1993c} Ma W X 1993 {\it J. Phys. A: Math. Gen.} {\bf 26} 2573
\bibitem{ChenZM} Chen D Y and Zhang H W 1991 {\it  J. Phys. A: Math. Gen.}
{\bf  24} 377
\end{thebibliography}
\end{document}